\documentclass[twocolumn,prb,showpacs,preprintnumbers,superscriptaddress,amsmath,amssymb]{revtex4}

\usepackage{graphicx}
\usepackage{dcolumn}
\usepackage{bm}

\begin{document}

\title{Multi-spin dynamics of the solid-state NMR Free Induction Decay}\

\author{H. Cho}
\affiliation{Department of Nuclear Engineering, Massachusetts Institute of Technology, Cambridge, MA 02139, USA}
\author{T. D. Ladd}
\affiliation{Department of Applied Physics, Stanford University, Palo Alto, CA 94305, USA}
\author{J. Baugh}
\affiliation{Institute for Quantum Computation, University of Waterloo, Waterloo, ON, Canada}
\author{D. G. Cory}
\affiliation{Department of Nuclear Engineering, Massachusetts Institute of Technology, Cambridge, MA 02139, USA}
\author{C. Ramanathan\footnote{Author to whom correspondence should be addressed. Electronic address:
sekhar@mit.edu}}
\affiliation{Department of Nuclear Engineering, Massachusetts Institute of Technology, Cambridge, MA 02139, USA}

\date{\today}

\begin{abstract}

 We present a new experimental investigation of the NMR
 free induction decay (FID) in a lattice of
spin-1/2 nuclei in a strong Zeeman field. Following a $\pi$/2 pulse, evolution under the
secular dipolar Hamiltonian preserves coherence number in the Zeeman
eigenbasis, but changes the number of correlated spins in the state.   The observed
signal is seen to decay as single-spin, single-quantum coherences evolve into multiple-spin 
coherences under the action of the dipolar Hamiltonian.  In order to
probe the multiple-spin dynamics during the FID, we measured the growth
of coherence orders in a basis other than the usual Zeeman eigenbasis.
This measurement provides the first direct experimental observation
of the growth of coherent multiple-spin correlations during the FID.
Experiments were performed with a cubic lattice of spins ($^{19}$F
in calcium fluoride) and a linear spin chain ($^{19}$F in
fluorapatite). It is seen that the geometrical arrangement of the
spins plays a significant role in the development of higher order
correlations.  The results are discussed in light of existing theoretical models.

\end{abstract}

\pacs{75.40.Gb, 76.60.-k, 82.56.-b}

\maketitle

Solid-state NMR is an ideal test-bed for studying coherent quantum
dynamics in a large Hilbert space. In this work, we experimentally
investigate the many-spin dynamics of the NMR free induction decay.
 The free induction decay (FID) is the response of the spin system
following a $\pi/2$ pulse. In a solid lattice of spin-1/2 nuclei in
a strong magnetic field, this evolution is dominated by the secular
dipolar Hamiltonian.  This is a classic problem in spin dynamics and has been well-studied
since the early days of NMR
\cite{Van_Vleck,Lowe,Abragam,Provotorov,Lado,Fine,Nevzorov,Nevzorov2,Engelsberg-2,Engelsberg,Munowitz_1}.
At the magnetic fields typically used in NMR experiments ($< 20$ T), 
the spin system is highly mixed at room temperature, and its equilibrium state
is represented by a thermal density matrix.  If the external Zeeman field is much 
stronger than the internal dipolar fields of the sample, the normalized deviation density
matrix in thermal equilibrium can be approximated as
\begin{equation}
\hat{\rho}(0)=-\sum_{j} {\hat{I}_{jz}} \; \; \; .
\end{equation}
Following a $\pi/2$ pulse, the spins  evolve under the secular dipolar Hamiltonian
\begin{equation}
\hat{H}_\mathrm{int}=\sum_{j<k}{D_{jk}\left\{\hat{I}_{jz}\hat{I}_{kz}-\frac{1}{4}\left(\hat{I}_{j+}\hat{I}_{k-}+\hat{I}_{j-}\hat{I}_{k+}\right)\right\}} \; \; \; .
\end{equation}
The strength of the dipolar coupling $D_{jk}$ between spins $j$ and
$k$ is given by
\begin{equation}
D_{jk}=\frac{\gamma^{2}\hbar^{2}}{{r}_{jk}^{3}}\left(1-3\cos^{2}\theta_{jk}\right) \; \; \;,
\end{equation}
where $\gamma$ is the gyromagnetic ratio, $r_{jk}$ is the distance
between spins $j$ and $k$, and $\theta_{jk}$ is the angle between
the external magnetic field and internuclear vector $\vec{r}_{jk}$.
Since the Hamiltonian is time-independent, the formal solution of the Liouville-von Neumann equation yields 
the density matrix of the spin system at time $t$ following the pulse as
 \begin{equation}
\hat{\rho}(t)=e^{-i \hat{H}_\mathrm{int}t / \hbar}\hat{\rho}(0)e^{i \hat{H}_\mathrm{int}t / \hbar}\;\;\; .
\end{equation}
An exact solution to this many-body problem has not been found, but the equation can be expanded
in a power series to examine the short time behavior of the system:
\begin{eqnarray}
\hat{\rho}(t)&=&\hat{\rho}(0)+\frac{i}{\hbar}t\left[\hat{\rho}(0),\hat{H}_\mathrm{int}\right] - \nonumber\\
&& \hspace{0.1in} \frac{t^2}{2{\hbar}^2}\left[\left[\hat{\rho}(0),\hat{H}_\mathrm{int}\right],\hat{H}_\mathrm{int}\right]+ \ldots  \label{fid0}
\end{eqnarray}
In an inductively detected NMR experiment (in which a coil is used
to measure the average magnetization), the observed signal is given
by
\begin{equation}
S(t)=\zeta \langle \hat{I}_{+} \rangle=\zeta \mathrm{Tr}\left\{\hat{I}_{+}\hat{\rho}(t)\right\} \;\;\; .\label {fid2}
\end{equation}
 where $\hat{I}_{+} = \sum_j (I_{jx} + i I_{jy})$ and $\zeta$ is a  
 proportionality constant.   The only terms in $\hat{\rho}(t)$ that yield a non-zero trace in the above equation and therefore contribute to the observed signal $S(t)$ are the single-spin angular momentum operators such as $\hat{I}_{j-}$, which are single-spin, single-quantum coherences.  Single quantum coherences are off-diagonal terms of the density matrix 
(in the Zeeman eigenbasis or the $z$-basis) connecting eigenstates with $\Delta m = \pm 1$ 
(corresponding to coherent superpositions of these eigenstates).  
Evaluating the commutators in Eq.~(\ref{fid0})
\begin{eqnarray}
\hat{\rho}(t)&=&-\frac{1}{2} \sum_{j}\left(\hat{I}_{j+}+\hat{I}_{j-}\right)+ \nonumber \\ && \frac{3}{2}it\sum_{jk}D_{jk}
 \left(-\hat{I}_{jz}\hat{I}_{k+}+\hat{I}_{jz}\hat{I}_{k-}\right)-   \label{fid1} \\ && 
 \frac{3}{4}t^{2}\sum_{jkl}D_{lk}D_{jk}\left(\hat{I}_{jz}\hat{I}_{lz}\hat{I}_{k+}+\hat{I}_{jz}\hat{I}_{lz}\hat{I}_{k-}\right)+\ldots \nonumber
\end{eqnarray}
Substituting Eq.~(\ref{fid1}) into Eq.~(\ref{fid2}), it can be seen
that the observable magnetization decays during the evolution under
$\hat{H}_\mathrm{int}$ because single-spin, single-quantum coherence
terms are transformed to unobservable multiple-spin, single-quantum
coherence terms by the higher-order nested commutators.   The $n$$^{th}$ 
 term in the expansion in Eq.~(\ref{fid0}) has $n$-spin correlations. 
 
There has been much theoretical effort to predict the
shape of the FID in calcium fluoride (CaF$_{2})$
\cite{Van_Vleck, Lowe,Abragam,Provotorov,Lado,Fine,Nevzorov,Nevzorov2}. Calcium
fluoride is a standard test system for spin dynamics as the $^{19}$F
(spin-1/2) nuclei are 100$\%$ abundant and form a simple cubic
lattice. The main goal has been to reproduce the decay and beat
pattern of the observed time domain NMR signal. For example,
Engelsberg and Lowe \cite{Engelsberg} measured up to 14 moments of
the FID in CaF$_{2}$, and these were found to be in good agreement
with theoretically calculated values for the 2nd to 8th moments. The
odd moments of the FID are zero, and the even moments are given by
\cite{Van_Vleck}
\begin{equation}
M_{2n}=\frac{(-1)^n}{\mathrm{Tr}\left\{\hat{I}_{x}^{2}\right\}}
\mathrm{Tr} \{ \underbrace{\left[\hat{H}_\mathrm{int},\left[\hat{H}_\mathrm{int},\left[\ldots,[\hat{H}_\mathrm{int},\hat{I}_{x}\right]\ldots\right]\right]}_{\mbox{2n times}}\hat{I}_{x}\} \;\;\; .
\label{moment1}
\end{equation}
Evaluating the nested commutators becomes increasingly challenging
and the higher-order moments are difficult to calculate. However, it
is these higher moments that characterize the many spin correlations
in the spin system. It can be seen that the 2$n$$^{th}$ moment arises
from the (2$n$+1)$^{th}$ term in the expansion in Eq.~(\ref{fid0}),
which creates up to  (2$n$+1) correlated spins. The main weakness of
the moment method lies in the fact that the most important
contribution to the value of the higher moments comes from the tails
of the FID, which are acquired with the lowest signal-to-noise ratio (SNR) in typical FID
measurements \cite{Abragam}.

 In this paper, we use a novel multiple-quantum NMR
 technique \cite{Sekhar} to study multiple-spin dynamics during the FID.
Standard multiple-quantum techniques \cite{Yen,Baum,Munowitz,Suter} encode coherence orders in the Zeeman 
 eigenbasis (or $z$-basis),
 but coherence numbers are conserved under the secular dipolar Hamiltonian in this basis \cite{footnote1}.  
 In our experiment we encode multiple-quantum coherences in the $x$-basis. The dipolar Hamiltonian in the $x$-basis is
\begin{eqnarray}
\hat{H}_\mathrm{int}^{x}&=&-\frac{1}{2}\sum_{j<k}D_{jk}\left\{\hat{I}_{jx}\hat{I}_{kx}-\frac{1}{4}\left(\hat{I}_{j+}^{x}\hat{I}_{k-}^{x}+\hat{I}_{j-}^{x}\hat{I}_{k+}^{x}\right)\right\} - \nonumber\\
&& \hspace*{0.4in} \frac{3}{8}\sum_{j<k}D_{jk}\left(\hat{I}_{j+}^{x}\hat{I}_{k+}^{x}+\hat{I}_{j-}^{x}\hat{I}_{k-}^{x}\right)\label{fid3} 
\end{eqnarray}
and no longer conserves coherence order in this basis (we use the superscript $x$ to denote that the raising and lowering operators are defined in the $x$-basis; these operators are otherwise assumed to be expressed in the $z$-basis).  The coherence orders are encoded by
 a collective rotation about the $x$-axis (which is the effective quantizing axis in this basis).
 Transforming the density matrix shown in Eq.~(\ref{fid1}) into the $x$-basis yields
\begin{eqnarray}
\hat{\rho}^x(t)&=&-\sum_{j}{\hat{I}_{jx}} -\frac{3}{4}it\sum_{jk}D_{ij}\left(\hat{I}_{j+}^{x}\hat{I}_{k+}^{x}-\hat{I}_{j-}^{x}\hat{I}_{k-}^{x}\right) + \nonumber \\ & & \hspace{0.2in} \frac{3}{8}t^{2}\sum_{jkl}D_{lk}D_{jk}\left(\hat{I}_{j+}^{x}\hat{I}_{l+}^{x}\hat{I}_{kx}
-\hat{I}_{j+}^{x}\hat{I}_{l-}^{x}\hat{I}_{kx} - \right. \nonumber \\ & & \hspace{0.4in} \left. \hat{I}_{j-}^{x}\hat{I}_{l+}^{x}\hat{I}_{kx}+\hat{I}_{j-}^{x}\hat{I}_{l-}^{x}\hat{I}_{kx}\right)+ \ldots   \label{fid4}
\end{eqnarray}
 From Eq.~(\ref{fid4}), it can be seen
that even order multiple quantum coherences are created in the $x$-basis.  
  It is possible to generate only odd order coherences using a $y$-basis
encoding for the same initial state. 
\begin{figure*}
\scalebox{0.6}{\includegraphics{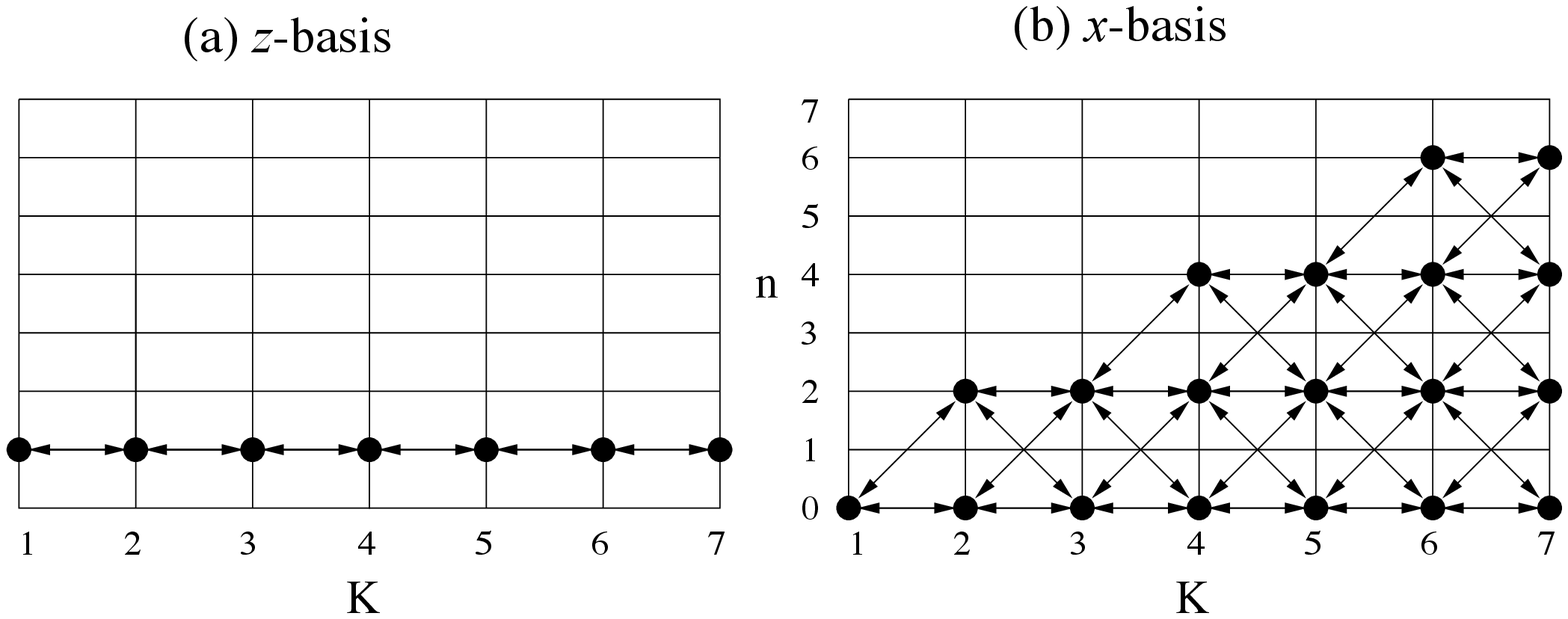}}
\caption{\label{Liouville} Projection of Liouville space onto the two-dimensional plane spanned by $K$ and $n$, showing the dynamics of the FID in (a) the Zeeman eigenbasis, and (b) the $x$-basis.  The arrows show the allowed paths in each case.}
\end{figure*}

It is useful to consider the dipolar evolution of this highly mixed state using the Liouville space formulation 
 for multiple-quantum dynamics  suggested previously \cite{Munowitz}. The
 density operator in Liouville space can represented as
 \begin{equation}
 \hat{\rho}(t)  = \sum_{K=0}^{N}\sum_{n=-K}^{K}\sum_p g_{Knp}(t) \hat{P}_{Knp}
 \label{eq:Liouville-1}
 \end{equation}
 where $\hat{P}_{Knp}$ represents a basis operator which is a 
 product of $K$ single-spin angular momentum operators, $n$ is the coherence order of the operator
  and $p$ is a label that identifies a particular configuration of spins having the 
 same $K$ and $n$.  
 The selection rules for the dipolar Hamiltonian in the Zeeman basis are given by
 \begin{equation}
 \Delta K = \pm 1 \: ,\: \: \: \: \Delta n = 0 \; \; \; .
 \end{equation}
 A projection of Liouville space onto the two-dimensional plane spanned by $K$ and $n$ is shown in Figure~\ref{Liouville}(a).  
 Following a $\pi$/2 pulse, the trajectory in the Zeeman basis is indicated by the arrows (only positive coherences are shown here;  the evolution is perfectly symmetric for negative $n$).  Increasing numbers of spins are correlated following evolution under the dipolar Hamiltonian, but the coherence number does not change.  Figure~\ref{Liouville}(b) shows the same evolution in the $x$-basis where the selection rules are
  \begin{equation}
 \Delta K = \pm 1 \: ,\: \: \: \: \Delta n = 0, \pm 2 \; \; \; .
 \end{equation}
Starting from an initial $I_x$ state ($K=1, n = 0$), only even order coherences are observed.  In this paper we characterize
the growth of these coherences.

\begin{figure}
\scalebox{0.45}{\includegraphics{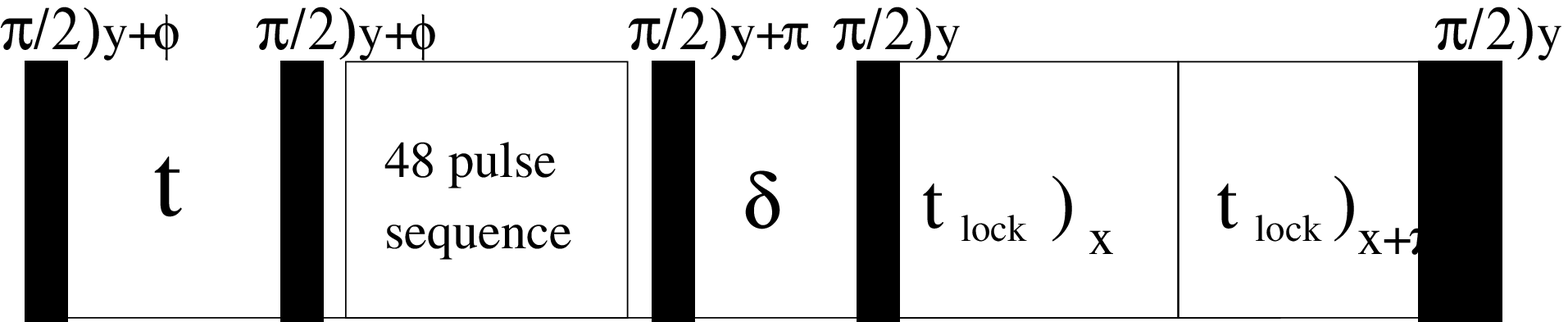}}
\caption{\label{pulsesequence}Pulse sequence used in this
experiment. $t$ is the evolution time under the secular dipolar
Hamiltonian. The 48-pulse sequence was used to suppress the
evolution of the internal Hamiltonian during the $x$-basis encoding
step. A magic-echo sequence was used to reverse the dipolar
evolution. A delay ($\delta$=10 $\mu$s) was inserted before the
magic echo to push the echo out, in order to minimize any pulse transients and receiver 
dead time effects. The duration of the spin locking time in the magic echo sequence
is $2t_{\mathrm{lock}}$ where $t_{\mathrm{lock}}=t+2\delta+3\mu$s.}
\end{figure}
The pulse sequence used in this experiment is shown in
Figure~\ref{pulsesequence}. After an initial $\pi/2$ pulse, multiple-spin, single-quantum 
states in the Zeeman basis are created during
evolution under the secular dipolar Hamiltonian as described in
Eq.~(\ref{fid1}). A $\phi \hat{I}_{x}$ rotation encodes coherence
orders in the $x$ basis, and a magic-echo sequence \cite{Rhim} is
used to refocus the multiple-spin terms back to observable single-spin, 
single-quantum coherence terms.  
The $\phi \hat{I}_x$ rotation is obtained by applying two $\pi/2$ pulses, with phases $y+\phi$ and $\bar{y}$, which results in 
the propagator $\exp (i\phi\hat{I}_z) \exp (i\phi\hat{I}_x)$.  The initial $\pi/2$ excitation pulse is also phase shifted by $\phi$ to cancel out the rotation about $\hat{I}_z$.  Since it is difficult to apply back to back $\pi / 2$ pulses without a delay between them (without introducing phase transients or allowing some dipolar evolution during the pulses), an evolution suspension sequence needs to be used in between the two $\pi/2$ pulses.  In this experiment we use a previously described 48-pulse evolution suspension  sequence \cite{Cory}.

\begin{figure}
\scalebox{0.5}{\includegraphics{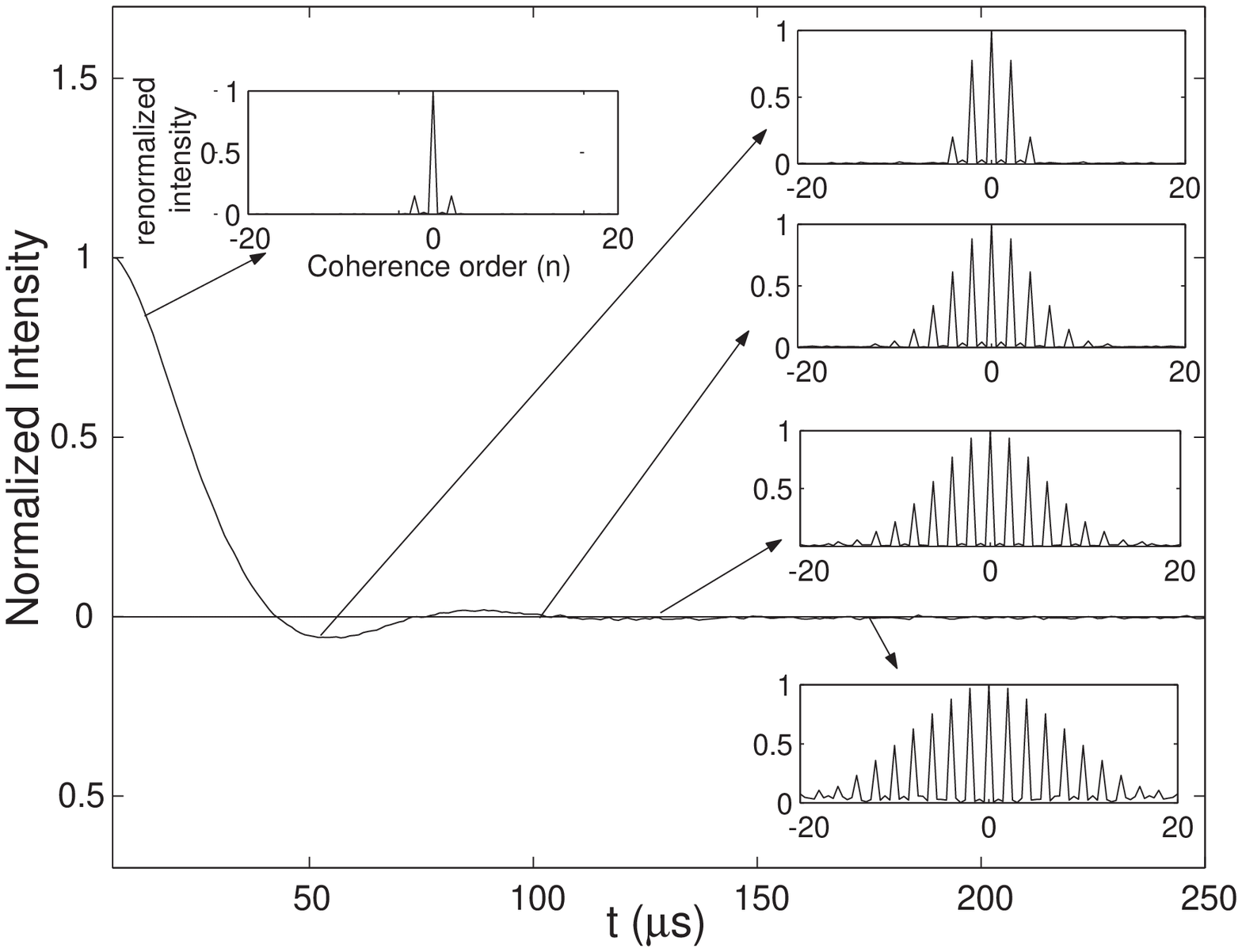}}
\caption{\label{xbasis} $x$-basis coherence order distribution at
various time points under the evolution of the secular dipolar
Hamiltonian in CaF$_{2}$. The multiple-quantum intensities have been re-normalized to set the intensity 
of the zero-quantum term to one in each case.}
\end{figure}

The experiments were performed at room temperature at 2.35 T
(94.2MHz, $^{19}$F), using a Bruker Avance spectrometer and home-built 
probe. The samples used were a 1 mm$^{3}$ single crystal of
CaF$_2$ with $T$$_{1}$ $\sim$ 7 s, and a crystal of fluorapatite
(FAp) with $T$$_{1}$ $\sim$ 200 ms.   The FAp crystal is a mineral crystal specimen from Durango, Mexico.
Such natural crystals are expected to have a number of interruptions in the spin chains as well as a variety of 
other defects \cite{Cho} as indicated by the short T$_1$.  All experiments were conducted
on resonance.  High power 0.5 $\mu$s $\pi$/2 pulses were used for the 48-pulse evolution
suspension sequence, while lower power 1.5 $\mu$s $\pi$/2 were used during the magic-echo
sequence as this sequence is more susceptible to phase transient errors.
 The phase $\phi$ was incremented
from 0 to 4$\pi$ with $\Delta\phi=\pi/32$ to encode up to 32
quantum coherences for every experiment. A fixed time point
corresponding to the maximum intensity signal was sampled for each
$\phi$ value and Fourier transformed with respect to $\phi$ to
obtain the coherence order distribution at each dipolar evolution
time $t$.

 Figure~\ref{xbasis} shows the coherence order distribution observed for CaF$_2$ at various time points
during the FID.  At short times, the maximum coherence order
($n_\mathrm{max}$) corresponds to the maximum number of correlated
spins ($K_\mathrm{max}$, i.e $g_{Knp} = 0$ for $K > K_{\mathrm{max}}$  ).  
At longer times, the maximum coherence order observed in the experiment sets the 
lower limit of the size of the spin correlation, since the SNR of higher order coherences might 
be too low to be observed.

\begin{figure}
\scalebox{0.5}{\includegraphics{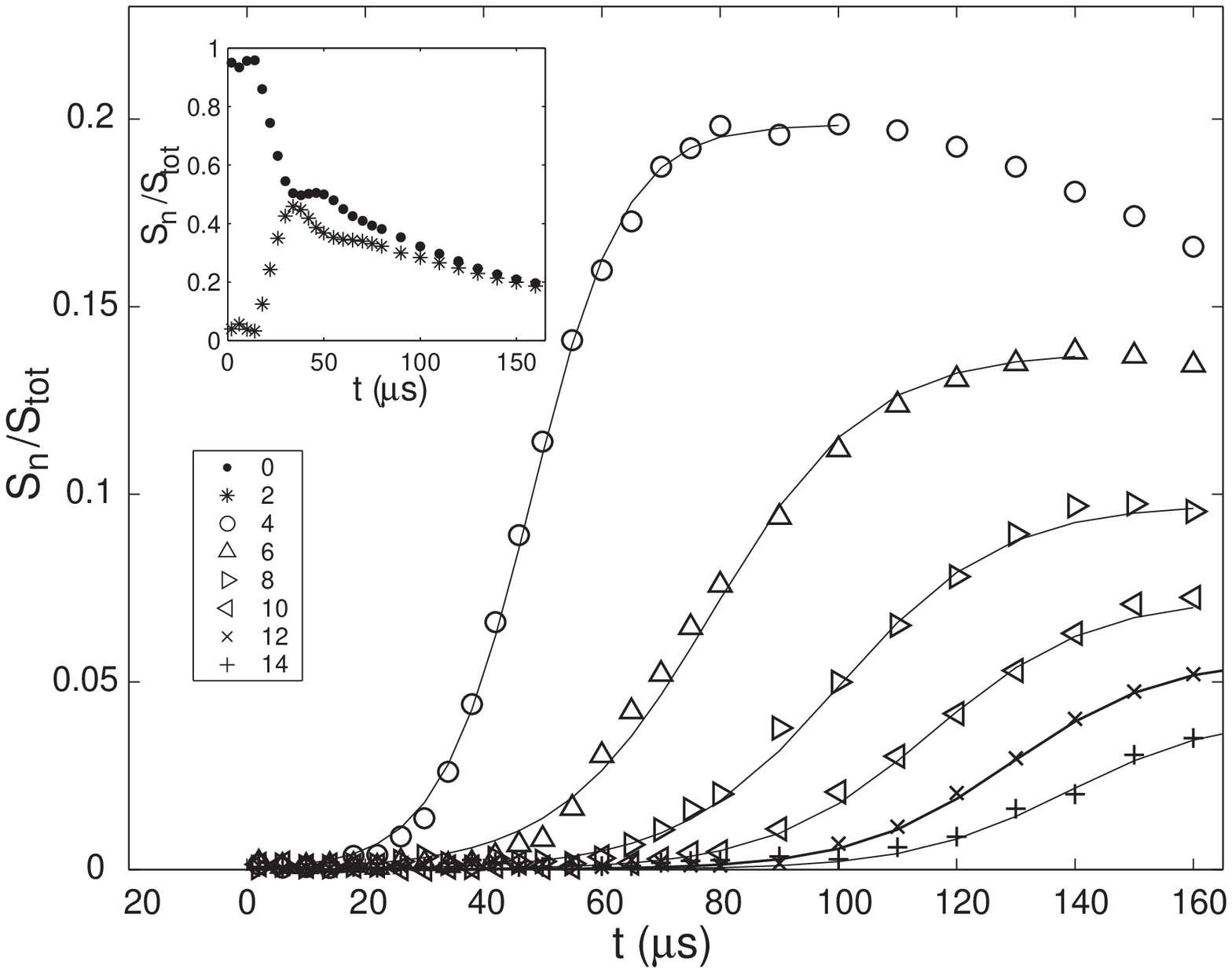}}
 \caption{\label{sigmoidal}The growth of multiple-spin correlations
during the FID, showing the sigmoidal fit to the initial growth data of
each $x$-basis coherence order ($n\geq4$), along $\sim$[110] direction in CaF$_{2}$.
 The inset shows the dynamics of $n = 0$ and $2$.  The intensity of the signal for each
coherence order $S_n$ is normalized with respect to the total
signal $S_{\mathrm{tot}}$, in order to compensate for imperfect refocusing of the dipolar interaction.}
 \end{figure}

Figure~\ref{sigmoidal} shows the growth of the different coherence
orders in CaF$_2$ during the FID. The inset shows the initial
oscillation between the zero and double quantum coherences at short
times (which corresponds to single and two spin correlations
respectively) due to the resolved nearest neighbor coupling at the
[100] and [110] directions.  This oscillation may be theoretically understood by considering the time 
development of an isolated pair of spins under the secular dipolar Hamiltonian (in the $x$-basis):
\begin{eqnarray}
\rho(t) & = & \frac{1}{2}\cos\left(\frac{3Dt}{2}\right)\left(I_{1x}+ I_{2x}\right) - \nonumber \\ & & \hspace{0.2in} \frac{i}{4}\sin\left(\frac{3Dt}{2}\right)\left(I_{1+}^xI_{2+}^x - I_{1-}^xI_{2-}^x\right) \label{eq:twospin}
\end{eqnarray}
where D is the strength of the pairwise coupling.
  In an extended spin system, this
oscillation is rapidly damped by leakage from isolated pairs to higher order correlations.
The higher order coherences ($n\geq$4) are seen to follow a sigmoidal growth curve.  
The higher order coherences develop later in time, and this progressive growth 
leads to a saturation of the intensities of the lower order coherences, 
consistent with the model in Figure 1.
As $t$ increases, imperfect refocusing of the 
dipolar evolution under the magic echo sequence results in a decay of the observed signal.  
In order to remove this decay the intensity for each coherence 
order is normalized with respect to the total signal measured at that evolution time.  

\begin{figure}
\scalebox{0.5}{\includegraphics{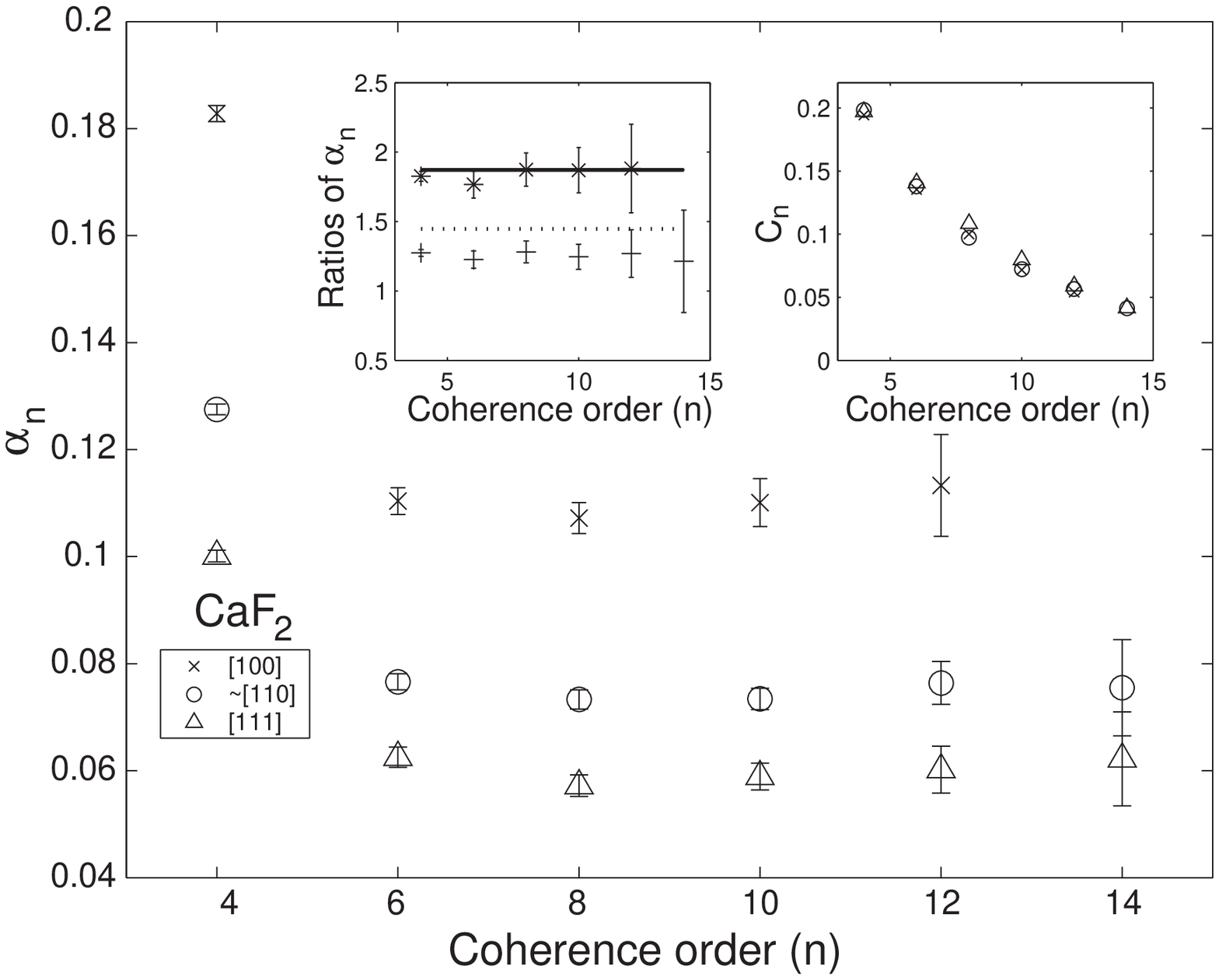}}
\caption{\label{onset}The values of $\alpha_{n}$ for different
orientations in CaF$_{2}$.  The left inset shows the ratios  $\alpha_{[100]}/\alpha_{[111]}$ (*) and 
$\alpha_{\sim[110]}/\alpha_{[111]}$(+) .  The calculated ratios of the mean dipolar coupling, obtained by averaging over 26
 nearest neighbors,  [100]/[111] (solid line) and $\sim$[110]/
[111] (dotted line) are also shown.  The right inset show the values of $C_{n}$'s
for different orientation in CaF$_{2}$}
\end{figure}

We have fit the initial growth of each coherence order to the following sigmoidal
function
\begin{equation}
S_{n}(t)=\frac{C_{n}}{1+e^{-\alpha_{n}(t-t_{n}^{\mathrm{onset}})}} \; \; \; .
\label{fid5} 
\end{equation}
In Figure~\ref{onset} we plot the variation of $\alpha_n$  and $C_n$
as a function of coherence order $n$.  The parameter $\alpha_{n}$ represents the underlying rate at
which the different coherence orders are transformed, and should be dominated by the  strength of 
the dipolar couplings involved.    It is seen that $\alpha_n$ does not vary with $n$, suggesting that the 
near-neighbor interactions dominate the dynamics here.  The mean dipolar coupling strength depends on the 
crystal orientation, so we measured  $\alpha_n$ with the crystal oriented along the [111], 
[110] and [100] directions with respect to the external field.  The mean dipolar coupling can be estimated by summing 
$|D_{1j}|$ over the 26 nearest neighbor spins for each crystal orientation.  For a simple cubic lattice, the ratio
of these means is 1:1.45:1.87 for [111], [110] and [100] in good agreement with the ratios of $\alpha_n$ shown in the figure.
The values of $C_{n}$ are seen to decrease as coherence number
increases, independently of crystal orientation.  This is expected as the total polarization is conserved, and the signal is progressively spread over increasingly larger regions of the system Hilbert space.
 
\begin{figure}
\scalebox{0.5}{\includegraphics{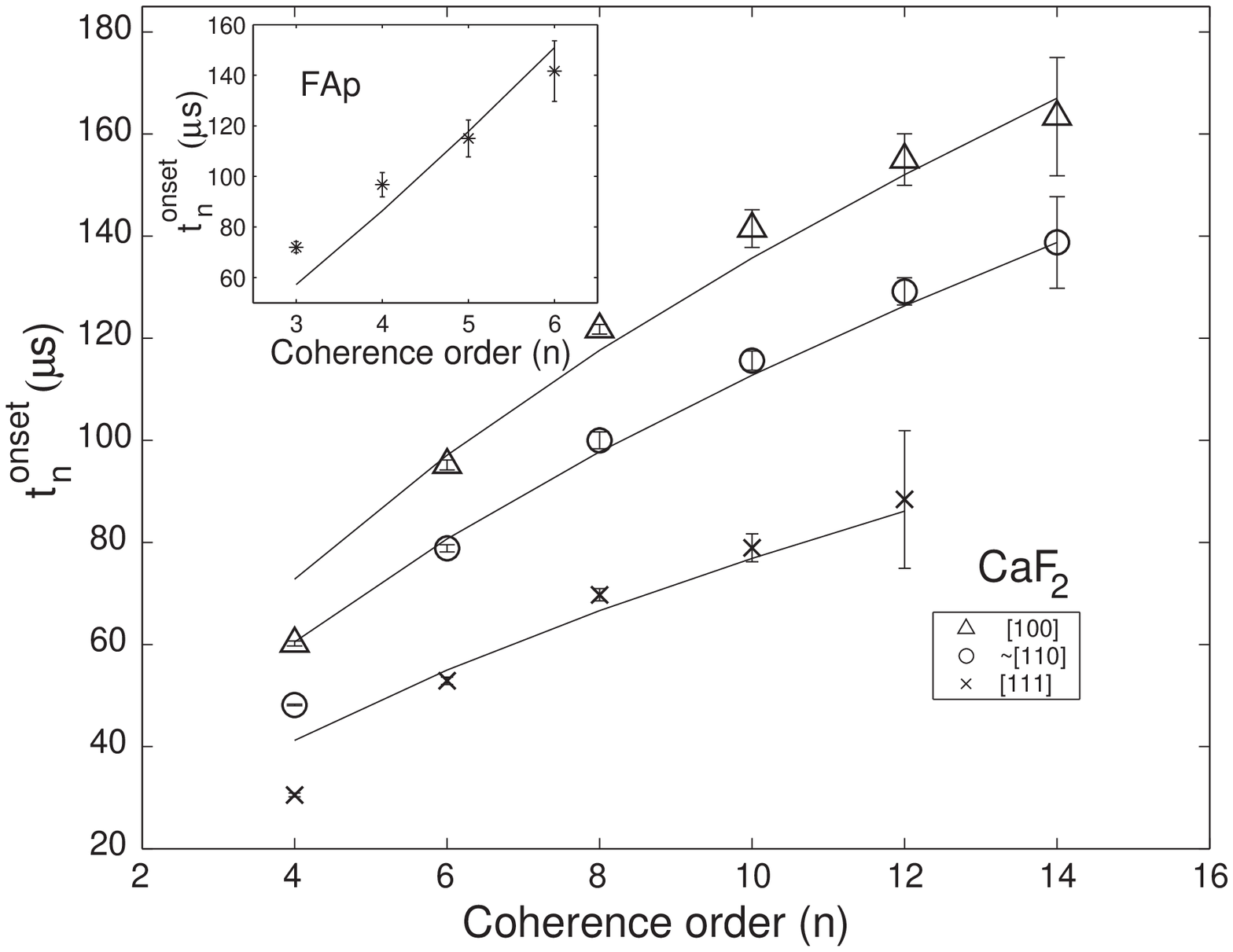}}
\caption{\label{onset2}Onset time of multiple-spin correlations for 
different crystal orientations in CaF$_{2}$.  The inset shows the onset times for the
FAp sample. Odd coherence orders for the FAp sample were obtained using a
$y$-basis encoding on the same initial state.  The continuous lines represent the best fits of Eq.~(\ref{eq:Thaddeus-1}) to the data, assuming that the same equation is valid for coherence number as well.}
\end{figure}

Figure~\ref{onset2} shows the onset time  $t_{n}^{\mathrm{onset}}$ 
of each of the $x$-basis coherences for different orientations of the crystal. 
 Physically, the onset time corresponds
to the time required for a specific coherence order to become
observable in the experiment.  While a first glance at Eq.~(\ref{fid1}) would seem
to suggest that higher-order correlations should develop as $t^n$,
it is the geometry of the spin system (the values of $D_{ij}$) which
ultimately determines the rate at which the spin correlations grow.
The onset times depend on the rate at which the correlations are spreading
through the spin system, which in turn depends on the value of the dipolar couplings.
Thus the rate is expected to be fastest (shorter onset time) with the crystal oriented along the [100] direction 
and slowest (longer onset time) for the crystal oriented along the [111] direction, in agreement with the experimental
data.   The variation of onset times with coherence order is also observed to depend on the dimensionality of the spin system.  The variation is sub-linear in the cubic CaF$_2$ system and displays an 
approximate $n^{2/3}$ dependence.  The inset in Figure~\ref{onset2} shows that the onset times obtained for FAp, the quasi-one-dimensional spin system, vary linearly with coherence number, in marked contrast to the results
from CaF$_2$.    

A variety of models have been proposed to describe the dynamics under a multiple-quantum Hamiltonian.
The most commonly used model involves a random walk among the components of the Liouville space basis set $\hat{P}_{Knp}$, 
subject  to the selection rules of the multiple quantum Hamiltonian \cite{Munowitz}.  The model replaces the Liouville-von Neumann equation by a set of coupled rate equations with exponential solutions 
\begin{equation}
\frac{d}{dt} \mathbf{g} = \mathbf{R \cdot g} \; \; \; ,
\end{equation}
where the vector {\bf g} containes the coefficents $g_{Knp}$.   All possible configurations that contribute to a particular coherence are assumed to be present in equal measure, and the resulting growth of the spin system is described by a hopping procedure between the allowed points on the lattice (shown in Figure~\ref{Liouville}). Under this assumption, the hopping rates are solely determined by the degeneracies of the coupled states.  The model thus eliminates any oscillatory solutions and precludes the possibility of quantum interference effects playing a role in the evolution.  The model also ignores the geometrical arrangement of the spin system and the specific distribution of dipolar coupling coefficients responsible for driving the evolution.  All spin systems display a universal growth kinetics, as long as the dynamics are scaled by a lattice parameter that represents the mean dipolar coupling strength of the  system, similar to the parameter $\alpha_n$ obtained above.  A numerical solution of the coupled equations for the multiple quantum evolution was observed to yield sigmoidal growth curves for the various coherence orders \cite{Munowitz}.  Limitations to this model have been discussed elsewhere \cite{Lacelle}.

Munowitz and Mehring \cite{Munowitz_1} used this model to numerically simulate the growth of multi-spin dynamics of the FID in a 21-spin system.   In order to track the development of correlations among the spins, they defined an induction time $t_K$ over which a particular $K$-spin coherence reaches half its maximum value.   This parameter is  very similar to the experimentally measured onset times of the different coherence orders described here.  Figure 7 of Ref.~\onlinecite{Munowitz_1} shows the variation of induction time with the number of correlated spins.  The numerical results show that the variation of the induction time is sub-linear for small numbers of correlated spins ($< 10$), in agreement with the experimental data.   For larger numbers of correlated spins, however, there is a marked deviation from the sub-linear behavior, as the effects of the finite system size (21 spins) begin to manifest themselves in the simulations.  The number of correlated spins would have to approach the number of spins in the sample ($\approx 10^{21}$) before such effects would be observed experimentally.  While providing some insight into the growth of the dynamics for the cubic spin systems, it is seen that the model does not do a very good job at describing the one-dimensional spin system.  It is in this situation that the spin geometry plays a dominant role.

Gleason and co-workers have proposed an alternate model to  describe the growth of spin correlations that emphasizes the geometrical ordering of spins \cite{Levy}.  By aggregating over the different configurations and coherence orders, i.e.\ summing over $n$ and $p$ in Eq.~(\ref{eq:Liouville-1}) the density operator $\hat{\rho}$ is expressed as a sum of terms with spin number $K$ and coefficients $g_K$,
\begin{equation}
\hat{\rho}(t) = \sum_{K=0}^{N} g_K(t) \hat{P}_K \: \: \:.
\end{equation}
where $g_K \hat{P}_{K}= \sum_{np} g_{Knp} \hat{P}_{Knp}$.
Essentially this model assumes a single effective $K$-spin operator that incorporates all the possible spin and spatial configurations of the $K$ spins.  The resulting model for spin propagation through a lattice yields a 
differential equation for the coefficients $g_K(t)$ of the form 
\begin{equation}
\frac{d}{dt} g_K = -i\left(W_{K-1}^\mathrm{f} g_{K-1} + W_{K+1}^\mathrm{r} g_{K+1}\right) \label{eq:Gleason2}
\end{equation}
where the rate constants $W^\mathrm{f}$ and $W^\mathrm{r}$ correspond to the forward and reverse rates respectively.  Under the assumption that the spatial grouping of the $K$ spins is continuous, and that only the nearest neighbor couplings are important, the forward rate (and equivalently the reverse rate) can be expressed as $W_K^\mathrm{f} \propto Dn_\mathrm{n}n_\mathrm{s}$, where $D$ is the strength of the nearest neighbor coupling, $n_\mathrm{s}$ is the number of spins on the surface of the spatial grouping, and $n_\mathrm{n}$ is the number of neighboring spins coupled to each spin.  New spins are added on the surface of the correlated spin cluster.
While $n_\mathrm{n}$ is a constant, the term $n_\mathrm{s}$ would differ significantly for spin systems of different dimensionalities, and can be expressed as $n_\mathrm{s} \propto  K^{1-1/d}$ where $d$ is the dimensionality of the spin system.   For a linear spin chain $d=1$ and $n_\mathrm{s}$ is independent of $K$, while for a cubic spin system, $d=3$ and $n_\mathrm{s} \propto K^{2/3}$.  While this model does not discuss coherence order, the dimensional dependence does agree with the experimental results, if the onset time characterizes the effective rate constant.  In the limit of large $K$, the rate constants $W_{K-1}^\mathrm{f} \approx W_{K+1}^\mathrm{r}$, and the coefficients $g_K(t)$ are approximately given by 
\begin{equation}
g_K(t) \propto i^{K-1}\left[\tanh\left(\beta t K^{-1/d}\right)\right]^K  \label{eq:thaddeus-2} \; \; \; ,
\end{equation}
where $\beta$ is proportional to the mean dipolar coupling.  With appropriate choice of normalization, the magnitude of $g_K(t)$ as given by Eq.~(\ref{eq:thaddeus-2}) strongly resembles the sigmoidal growth characteristics of the multiple quantum coherence intensities shown in Figure \ref{sigmoidal}.   An onset time can be obtained from Eq.~(\ref{eq:thaddeus-2})  by setting $[\tanh(\beta t K^{-1/d})]^K = 1/2$, yielding
\begin{equation}
t_{1/2} = \frac{\eta}{\beta} K^{1/d} \mathrm{arctanh} \left(2^{-1/K}\right) \; \; \; ,
\label{eq:Thaddeus-1}
\end{equation}
where $\eta$ is a constant scaling factor.  Figure \ref{onset2} shows the best fit of Eq.~(\ref{eq:Thaddeus-1}) to the experimental data, assuming that the model holds true for coherence number as well.   It is seen that there is excellent agreement at larger values of $n$ for the cubic CaF$_2$ system.  The values of $\eta/\beta$ obtained from the fit are 37.47 in the [111] direction, 31.13 in the [110] direction and 21.22 in the [001] direction.  Their inverses are in the ratio 1:1.46:1.76  for [111]:[110]:[100] which is in excellent agreement with the theoretically calculated values shown earlier.  For the linear FAp system,  Eq.~(\ref{eq:Thaddeus-1}), which is linear for large $K$, is observed to be weakly superlinear at these small values of $K$ \cite{note2}.

The constancy of $\alpha_n$ in the sigmoidal plots in Figure~\ref{sigmoidal} and the good agreement observed  between the observed onset times and Eq.~(\ref{eq:Thaddeus-1}) indicate that the spin dynamics are dominated by the nearest-neighbor interactions in this regime.  This is not surprising, as we are still operating in the short time regime.  Higher-order spin processes, if significant, would be expected to manifest themselves at later times, leading to a deviation from the simple model behavior described above.

In conclusion, we have presented a new experimental method to characterize
 multi-spin dynamics in solid-state NMR free induction decay.  The
 initial creation of coherences were observed to follow a sigmoidal growth curve, with
the onset times characterizing the dynamics of the spin system.
These dynamics in turn were critically dependent on the geometrical
arrangement of the spins as expected. 

 The authors thank D. Greenbaum, P. Capperallo and T. S. Mahesh for helpful
discussions, and the NSF, ARO and DARPA DSO
for financial support.


\begin{thebibliography}{00}
\bibitem{footnote1}
The conservation of coherence order under the dipolar Hamiltonian is a consequence of the truncation of the non-secular components of the dipolar interaction in a strong Zeeman field.  The non-secular components can change coherence orders by $\pm1$ or $\pm2$.
\bibitem{note2}
Examining Eq.~(\ref{eq:Gleason2}), we see that during the initial growth of $g_K$ is dominated by $W_{K-1}^\mathrm{f}$ as $g_{K+1} = 0$.   Thus the incremental time to go from a $(K-1)$-spin system to a $K$-spin system
is inversely proportional to $W_{K-1}^\mathrm{f}$.    Applying this recursively, we obtain a general expression for the onset time $t_K = \eta H_{K-1}^{1-1/d}$, where $\eta$ is a constant and $H_n^r$ is a generalized Harmonic number of order $r$.
The fits obtained with this equation are also in good agreement with the experimental data, and appear to be functionally identical to the Eq.~(\ref{eq:Thaddeus-1}) in the range of large $K$.
\end{thebibliography}
\end{document}